\def\outtype{1}
\ifnum\outtype=1
\documentstyle[12pt,aaspp4,astrobib]{article}
\else
\documentstyle[12pt,aasms4,astrobib]{article}
\fi
\begin{document}
\title{An Instability in the Radiative Ionization of Atomic
  Hydrogen/Helium Gas
\ifnum\outtype=1
\footnote{Accepted by the Astrophysical Journal.}
\fi
}
\author{Ethan Bradford\footnote{To whom correspondence should be
    addressed, preferably at ethanb@phys.washington.edu.\ifnum\outtype=1
 All comments and suggestions are welcome.
\fi
}
 and Craig Hogan}
\affil{Dept. of Physics, FM-15\\University of Washington, Seattle, WA 98195}
\authoraddr{}

\begin{abstract}

We show that 
the process of photoionizing 
a gas of atomic hydrogen and helium by line radiation
whose energy is slightly above the helium single-ionization threshold
is unstable if
the helium fraction by number is less than approximately one half.
However, in the two 
scenarios we consider here, based on the
Decaying Dark Matter (DDM) model of cosmological reionization,
there is no significant growth.
In the first scenario we consider ionization and recombination to be
approximately in equilibrium.  This is relevant to high photon flux
rates and early reionization, but in that case the heating is balanced
by Compton cooling, which is very stabilizing.
In the second scenario we ignore recombination.  This is relevant to
low photon flux
rates or to the last stage of the reionization.
In that case there is too little growth on a cosmological time scale
to be significant.

\end{abstract}


\section{Introduction}

The hydrogen in the 
intergalactic medium (IGM) is highly ionized and has been at least since
$z=4.3$ \cite{Steidel:Sargent,GP4.7}.
Recent observations indicate that the helium in the IGM is mostly 
at least singly ionized, at least at large redshifts
\cite{Reimers&Vogel:HeGP,Jakobsen:HeGP,Miralda-Escude:HeGP}.
Though the IGM might be ionized mainly by
quasars \cite{Meiksin:Madau:Q2,Madau:Q1} or an
early generation of stars \cite{FK:Reheating},
additional ionization by photons emitted by decaying
dark matter (DDM) has many appealing
features \cite{Flannery:Press,Sciama:overview,Sciama:ddm_book,Scott:Rees:Sciama}.

In particular, if a generation of neutrinos has
a mass sufficient to close the universe, and if such a particle decays
into a much lighter particle and a photon, the photon's
energy will be near the ionization
energies of hydrogen and helium.
Specifically,
\begin{equation}
E_\gamma\approx{1 \over 2} m_\nu = {1 \over 2} 91.5 h^2 \,{\rm eV}
\end{equation}
where $h$ is the Hubble constant in units of 100 km/s/Mpc.
Thus, for $h > .55$, the decay photons would have sufficient energy to
ionize hydrogen and for $h > .74$, they could (singly) ionize helium.

The decay photons would also provide a source of out-of-equilibrium energy
which might lead to small-scale structure formation.
\citeN{Hogan:ion_inst}
proposed a mechanism of small-scale structure formation
through the ionization of a
pure-hydrogen plasma, but \citeN{me:no_ion_inst}
showed that this mechanism does not work.
We have continued to seek possible instabilities of the system, since
a neutrino-dominated universe requires a
non-gravitational mechanism for
spawning structure on galactic scale and below
and there is no fundamental reason why this system should be stable.
The free energy per atom is sufficient for many e-foldings of
instability growth ($E_\gamma \gg k T$), and the scale of instability
can be comparable to proto-galactic structure.
A hydrodynamical instability
would mimic many of the desirable features
found in the gas component of a Cold Dark Matter (CDM) model, as described 
for example in \citeN{Miralda-Escude:Mini-Halo}.

\subsection{Helium-Ionization Instability Mechanism}

\citeN{Meiksin} suggested that there is an instability if DDM photons
have sufficient energy to ionize helium.  The instability is thermal,
i.e. it is based on low-density regions being preferentially heated.
Low-density regions receive more heat because they have a larger ratio
of neutral hydrogen to neutral helium and a hydrogen ionization
deposits more energy in the plasma than a helium ionization (because
there is a larger difference between the photon energy and the
ionization threshold).  The ratio of neutral hydrogen to neutral
helium is larger in low density regions because they are more
ionized and the higher ionization cross-section of helium (for photons
just above its ionization threshold) ensures that it will absorb photons
out of proportion to its abundance.

The reason that low-density regions are more ionized is different in
the two scenarios considered below.  
When recombination equilibrium holds, it is due to
the fact that recombination happens in proportion to the square of the
density (because a positive ion and an electron need to find each
other) while the ionization rate is proportional to the density (the
rate of production of photons is assumed to be unrelated to the
density).
When recombination equilibrium doesn't hold, the low density regions
are more ionized because there are
more photons per atom.

Though any source of uniformly
distributed photons will do for this instability, we concentrate on
decaying neutrinos and parameterize the photon production rate by the
neutrino lifetime in units of $10^{24}$ seconds, $\tau_{24}$.

\section{Perturbation Expansion and Matter Evolution Equations}

We expand to first order in small perturbations about background
values, and
expand the perturbations in a Fourier series.
We normalize some perturbations by the unperturbed value, e.g. for
the number density, $n$,
\begin{equation}
n=n_0\left(1+\delta \ln n\; e^{i\vec{k} \bullet \vec{x}}\right)
\end{equation}
The zero superscript denotes the unperturbed variable; it will be
dropped in subsequent formulas for simplicity.

The evolution equation for the internal energy density, $u$, is
\begin{equation}
\label{udot1}
{\partial \over \partial t}u=\gamma u {\partial \over \partial t}\ln n+{\cal P}
\end{equation}
where $n$ is the number density, $\gamma$ is the adiabatic
expansion coefficient (5/3 for a monatomic gas), and 
${\cal P}$ is the heating and cooling
term.\footnote{Table~1 lists the symbols used in this article.}
This equation linearizes to
\begin{equation}
  \label{udot-l}
  {\cal P} \delta \ln u + u {\partial \over \partial t} \delta \ln u = 
  \gamma u {\partial \over \partial t} \delta \ln n + \delta {\cal P} +
  {\cal P}_u \delta \ln u  
\end{equation}

The relationship between density and pressure (or internal energy)
perturbations is
(see \cite{Peebles:LSS})
\begin{equation}
  \label{w:k}
  {\partial \over \partial t} {\partial \over \partial t} \delta \ln n +
  2 H {\partial \over \partial t} \delta \ln n = 
  - {k^2 \over a^2} {v_s^2 \over \gamma} \delta \ln u
\end{equation}
where $k$ is the comoving wave number,
$H$ is the Hubble constant,
$a$ is the cosmic scale factor and is equal to $1/(1+z)$,
and $v_s$ is the speed of sound (remember that the pressure
is just $\gamma-1$ times the internal energy density).

Here
\begin{equation}
  {v_s^2 \over \gamma} = {p_0 \over \rho_0} =
  {(1+\chi )T \over m_{proton}(1+3\mu)} =
  {(\gamma-1) u \over n m_{proton}(1+3\mu)}
\end{equation}
where $\mu$ is the fraction by number of helium nuclei and $\chi$ is
the ratio of free electrons to total hydrogen and helium nuclei.
Note that ${m_{\rm proton}(1+3\mu) \over (1+\chi )}$ is the average mass of a
plasma particle (since $m_{\rm electron} \ll m_{\rm proton}$).

Combining eqs.~\ref{udot-l} and \ref{w:k} gives
\begin{equation}
  \label{norec-1}
  {\partial \over \partial t} {\partial \over \partial t}
  {\partial \over \partial t} \delta \ln n +
  a_2
  {\partial \over \partial t} {\partial \over \partial t} \delta \ln n +
  a_1
  {\partial \over \partial t} \delta \ln n +
  {v_s^2 k^2 \over a^2 \gamma u} \delta {\cal P} = 0
\end{equation}
\begin{equation}
  a_1 = 2 H \left({{\cal P} \over u}-{\cal P}_u\right) +
         H^2 + v_s^2 {k^2 \over a^2}
  \label{a_1}
\end{equation}
\begin{equation}
  a_2 = {{\cal P} \over u}-{\cal P}_u + 4 H
  \label{a_2}
\end{equation}
Since $a_1$ and $a_2$ are positive
(note that $P_u$ is zero for ionization heating and is negative
for Compton cooling), we will have
an instability if
${\delta {\cal P} \over \delta \ln n} < 0$,
which we will show to hold
for the ionization instability considered alone, though it doesn't
hold when the ionization heating is balanced by Compton cooling.

Since $a_1$ and $a_2$ are positive, they lead only to damping of
growth.  We ignore them from now on, which can only be generous
to the instability.

The analysis to this point has been general to any sort of heating.
The form of $\delta \ln {\cal P}$ will be different for the two
instability scenarios here, though in both ${\cal P}$ depends directly
only on $\chi$ and the relationship between $n$ and $\chi$ is used to
get the variation of ${\cal P}$ with $n$.
For the ``recombination equilibrium'' scenario, we will find 
$\delta {\cal P}$ in terms of $\delta \ln n$ and the equation will be
third order;
for the ``no recombination'' scenario we will have
$\delta {\cal P}$ in terms of $\int \delta \ln n d t$ and the equation
will be fourth order.

\section{Scenario in which Recombination and Cooling are Important}
\label{Eq_sec}


\subsection{Photon Energy Levels}

At the heart of the ionization instability mechanism is the fact that
a helium ionization deposits more energy than a hydrogen ionization.
However, a significant part of that difference is lost when the helium
atom recombines, emitting photons which might then ionize a
hydrogen atom.

A direct recombination to ground will yield a photon
at the ionization energy of He, which we label $E_1$;
if this photon ionizes a hydrogen atom, no energy will have been
lost in the helium ionization.
A recombination to an excited state of helium will emit a low-energy
photon which can't ionize hydrogen, so its energy is lost.
All such recombinations
cascade (emitting other low-energy photons) to one of three states,
which then decays directly to ground, usually emitting a photon which
can ionize hydrogen (so we need to keep track of it).
The $2 {}\,^3\!P$ triplet state and the $2 {}\,^1\!P$ singlet
decay by emitting one photon at energy $E_2$ or $E_3$ respectively.
Helium atoms which end up 
in the He $2 {}\,^1\!S$ state
can only decay to ground by the emission of two photons.
We label the $2 {}\,^1\!S$ state energy $E_4$ and
the average energy of the two-photon-decay photons which can ionize
hydrogen $E_5$ (see eq.~(\ref{E5})).
Finally, we label the the primary photon energy
(the energy of the decay photons)
$E_0$.

On average
$p_1=0.554$ hydrogen ionizations
result from each two-photon decay
of the $2 {}\,^1\!S$ state (eq.~(\ref{p1}));
all other helium recombinations lead (in the optically thick limit) to a
hydrogen ionization.

See \citeN{Osterbrock:1} for more background on these
radiative processes.

\subsection{Ionization and Recombination Equations}

The heat energy absorbed will be proportional to the numbers of 
photons in each level times the absorption cross section for those photons:
\begin{eqnarray}
  {\cal P}=&n_{\gamma 0}[(E_0-{\rm Ry})\sigma_{\rm H 0}n_{\rm H\,I}+(E_0-
  E_1)\sigma_{\rm He 0}n_{\rm He\,I}]+n_{\gamma 1}( E_1-{\rm Ry})\sigma_{\rm
    H 1}n_{\rm H\,I}+\nonumber\\ 
  &n_{\gamma 2}(E_2-{\rm Ry})\sigma_{\rm H 2}n_{\rm H\,I}+n_{\gamma 3}(
  E_3-{\rm Ry})\sigma_{\rm H 3}n_{\rm H\,I}+n_{\gamma 5}(E_5-{\rm Ry})\sigma_{\rm
    H 5}n_{\rm H\,I}.
  \label{P}
\end{eqnarray}
$n_{\rm H\,I}$, $n_{\rm H\,II}$, $n_{\rm He\,I}$, $n_{\rm He\,II}$ are the
number densities of neutral and singly ionized hydrogen and helium atoms.
$n_{\gamma 0}$, $n_{\gamma 1}$, etc., are the number densities of
photons with energy $E_0$, $E_1$, etc.  ${\rm Ry}$ is one Rydberg, the
hydrogen ionization energy.
The sigmas are
the absorption cross sections for the given substances with
photons of the indicated energy, e.g. $\sigma_{\rm He 1}$ is the
cross section for the ionization of helium by photons with energy $E_1$.

The following equations express the assumption that the creation and
destruction of photons
at each level are in equilibrium:
\begin{eqnarray}
\Re&=&(\sigma_{\rm He 0}n_{\rm He\,I}+\sigma_{\rm H 0}n_{\rm
  H\,I})n_{\gamma 0}\nonumber\\
\alpha_{{\rm He} 1}n_{\rm He\,II}n_e&=&(\sigma_{\rm He 1}n_{\rm
  He\,I}+\sigma_{\rm H 1}n_{\rm H\,I})n_{\gamma 1}\nonumber\\
\alpha_{{\rm He} 2 {}\,^3\!P}n_{\rm He\,II}n_e&=&\sigma_{\rm H
  2}n_{\rm H\,I}n_{\gamma 2}\nonumber\\
\alpha_{{\rm He} 2 {}\,^1\!P}n_{\rm He\,II}n_e &=&\sigma_{\rm H
  3}n_{\rm H\,I}n_{\gamma 3}\nonumber\\
\alpha_{{\rm He} 2 {}\,^1\!S}n_{\rm He\,II}n_e p_1&=&\sigma_{\rm H 5}n_{\rm H\,I}n_{\gamma 5}
  \label{photon:balance}
\end{eqnarray}
$\Re$ is the rate of production of primary photons per unit volume
(in the decaying neutrino model it is $n_\nu / \tau_\nu$).
The $\alpha$'s are the net recombination coefficients into various
states, as defined in Table~1.
Direct recombination of hydrogen to ground is ignored since (in the
optically thick limit) the photon produced immediately ionizes another
neutral hydrogen atom.

The following equations say that the creation and destruction of each
species of ion are in equilibrium:
\begin{eqnarray}
\alpha_{{\rm H} 2}n_{\rm H\,II}n_e &=&(\sigma_{\rm H 0}n_{\gamma 0}+\sigma_{\rm H 1}n_{\gamma 1}+\sigma_{\rm H 2}n_{\gamma 2}+\sigma_{\rm H 3}n_{\gamma 3}+\sigma_{\rm H 5}n_{\gamma 5}) n_{\rm H\,I}\nonumber\\
\alpha_{\rm He}n_{\rm He\,II}n_e&=&(\sigma_{\rm He 0}n_{\gamma 0}+\sigma_{\rm He 1}n_{\gamma 1}) n_{\rm He\,I}
  \label{ion:balance}
\end{eqnarray}

We also use the following relationships (from
\citeN{Osterbrock:1}) between recombination coefficients:
\begin{eqnarray}
 \alpha_{\rm He}&=& \alpha_{{\rm He} 1}+ \alpha_{{\rm He} 2}\nonumber\\
 \alpha_{{\rm He} 2 {}\,^3\!P}&=&{3 \over 4}\alpha_{{\rm He} 2}\nonumber\\
  \alpha_{{\rm He} 2 {}\,^1\!P}&=& {2 \over 3}
    \left({1 \over 4}\alpha_{{\rm He} 2}\right)\nonumber\\
 \alpha_{{\rm He} 2 {}\,^1\!S}&=& {1 \over 3}
    \left({1 \over 4}\alpha_{{\rm He} 2}\right)
  \label{recomb-rel}
\end{eqnarray}

\subsection{Analysis}
Combining eqs.~(\ref{photon:balance}), (\ref{ion:balance}) and (\ref{recomb-rel}) gives
the density and hydrogen ionization coefficient in terms of
the helium ionization coefficient:
\begin{equation}
  \chi_{\rm H}={(1-\mu ) \sigma_{\rm H 1}\alpha_{\rm He}+{1 \over 12}( p_1+11)\sigma_{\rm He 1}\mu (1- \chi_{\rm He}) \alpha_{{\rm He} 2}\over (1-\mu )(\sigma_{\rm He 1}(1- \chi_{\rm He}) \alpha_{{\rm H} 2}+\sigma_{\rm H 1}\chi_{\rm He}\alpha_{\rm He})}\chi_{\rm He}
  \label{chi_h1}
\end{equation}
\begin{equation}
  n^2={\sigma_{\rm He 1}\Re (1- \chi_{\rm He}) \over ((1-\mu )\sigma_{\rm H 1}(1- \chi_{\rm H}) \alpha_{\rm He}+\sigma_{\rm He 1}\mu (1- \chi_{\rm He}) \alpha_{{\rm He} 2})((1-\mu ) \chi_{\rm H}+\mu  \chi_{\rm He})\chi_{\rm He}}
  \label{n_vs_chi}
\end{equation}

Applying those relations to the formula for ${\cal P}$ (eq.~(\ref{P})) gives
\begin{eqnarray}
{\cal P}&=&\Re \left({\Delta E}-{\sigma_{\rm He 1}\mu  \alpha_{{\rm
    He} 2}\over\sigma_{\rm He 1}\mu  \alpha_{{\rm He} 2}+\sigma_{\rm
    H 1}(1-\mu) \alpha_{\rm He}}E_{\rm Loss}\right) {1-A \chi_{\rm
    He}\over {1-B\chi_{\rm He}}}\\
A&=&{({\Delta E}-E_{\rm Loss})\sigma_{\rm He 1}\alpha_{{\rm H}
    2}+\left({E_{\rm Loss}-{1\over 12}(1-p_1){\Delta
      E}}\right)\sigma_{\rm H 1}\alpha_{\rm He}
 \over
    ({\Delta E}-E_{\rm Loss})\sigma_{\rm He 1}\mu\alpha_{{\rm He}
      2}+{\Delta E}\sigma_{\rm H 1}(1-\mu)\alpha_{\rm He}
 }
 \mu
{\alpha_{{\rm He} 2} \over \alpha_{{\rm H} 2}}\\
B&=& {\sigma_{\rm He 1}\alpha_{{\rm H} 2}-{1\over
    12}(1-p_1)\sigma_{\rm H 1}\alpha_{\rm He}
\over
 \sigma_{\rm H 1} (1-\mu)\alpha_{\rm He}+\sigma_{\rm He
   1}\mu\alpha_{{\rm He} 2}}
  \mu{\alpha_{{\rm He} 2}\over  \alpha_{{\rm H} 2}}\\
\end{eqnarray}
$\Delta E$ is the difference between $E_0$ and a Rydberg and
$E_{\rm Loss}$
is the difference in the energy deposited when a primary photon
ionizes a hydrogen atom instead of a helium atom:
\begin{equation}
 E_{\rm Loss}=E_1-{1\over 12}( E_5 p_1+2 E_3+9 E_2+{\rm Ry} (1-
 p_1))=4.97\,{\rm eV}.
\end{equation}

To get positive feedback in the heating, we need the heating to go
down with increasing density
and hence to go up with increasing ionization.  For that we need $A<B$ or
\begin{equation}
  \sigma_{\rm He 1}(1-\mu ) \alpha_{{\rm H} 2} >
  \sigma_{\rm H 1}(1-\mu ) \alpha_{\rm He} +
    {1 \over 12}( p_1+11)\sigma_{\rm He 1} \mu \alpha_{{\rm He} 2}
    \label{growth-condition}
\end{equation}
which holds since $\mu \lesssim {1 \over 2}$ (i.e. there is more hydrogen
than helium),
$\sigma_{\rm H 1}\ll\sigma_{\rm He 1}$ 
(i.e. the absorption cross-section for helium is larger at this energy),
and ${2 \over 3} \alpha_{\rm He} \approx \alpha_{{\rm He} 2} \approx
\alpha_{{\rm H} 2}$ (i.e. all the recombination coefficients are comparable). 

Filling in the numeric values given in Table~1 
and taking $E_0=E_1$ (that is, ionization photons right at the helium
ionization threshold, which is most generous to the instability),
we get
${\cal P}$ and $n^2$ in terms of the variable $\chi_{\rm He}$
and the parameter $\tau_{24}$:
\begin{eqnarray}
{\cal P}&=&1.12\times{10}^{-21}{{\rm eV} {z'}^3 \over {\rm cm}^3 s}
  {{1-0.180\chi_{\rm He}}
  \over 1-0.246 \chi_{\rm He}}
\label{P-num}\\
   n^2&=&2.29\times{10}^{-9}
   { 1 \over \tau_{24} {\rm cm}^6}
   {{\left(1-0.736 \chi_{\rm He}\right)}^2
   \over
   \left(1-0.243 \chi_{\rm He}\right)
   \left(1-0.326 \chi_{\rm He}\right){\chi_{\rm He}}^2}
\label{n-num}
\end{eqnarray}
where $z'=1+z$.
When varied these give
\begin{equation}
\delta {\cal P}=0.0662 \, {\cal P}
  {1  \over (1-0.180 \chi_{\rm He}) (1-0.246 \chi_{\rm He})}
  \delta \chi_{\rm He}
\end{equation}
(note the small coefficient in this equation, indicating the weakness
of this instability)
and
\begin{equation}
  \delta \chi_{\rm He}={\delta n  \over n}
  {{\left(1-0.246 \chi_{\rm He}\right)}^2
    {\left(1-0.351 \chi_{\rm He}\right)}^2
    {\left(1-0.736 \chi_{\rm He}\right)}^2
    {\chi_{\rm He}}
   \over
    (1-0.600\chi_{\rm He}+0.266{\chi_{\rm He}}^2)
   \left(1-0.243 \chi_{\rm He}\right)
   \left(1-0.295 \chi_{\rm He}\right)
   \left(1-0.326 \chi_{\rm He}\right)
   \left(1-0.771 \chi_{\rm He}\right)}
\end{equation}

Defining the coefficient multiplying $\delta \ln n$ in
eq.~\ref{norec-1} as
\begin{equation}
  a_0={v_s^2 k^2 \over a^2 \gamma u} {\cal P}_n
\end{equation}
and using $a^{-2}=(H/H_0)^3$ we get
\begin{equation}
  a_0= -{2.56
    {(1-0.35\chi_{\rm He})}^2{(1-0.74\chi_{\rm He})}^2
    \chi_{\rm He}
    ({\rm Mpc} \, k)^2 H^3
    \over
    (1-0.60\chi_{\rm He}+0.23{\chi_{\rm He}}^2)
    (1-0.24\chi_{\rm He})
    (1-0.29\chi_{\rm He})
    (1-0.33\chi_{\rm He})
    (1-0.77\chi_{\rm He})
    {z'}^{5/2} \tau_{24}}
\end{equation}

We can use eq.~\ref{n-num} and the known background value for the density,
$n=1.04\times10^{-7} {z'^3 \over cm^3}$
(this assumes $\eta=3\times10^{10}$) to get the background ionization
for a given redshift:
\begin{equation}
\label{tau-z}
\tau_{24}{z'}^3=4.06\times 10^{5}{{(1-0.78\chi_{\rm He})}^2\over{\chi_{\rm He}}^2(1-0.27\chi_{\rm He})(1-0.37\chi_{\rm He})}
\label{eq_tau}
\end{equation}
Using this equation to eliminate the redshift gives
\begin{equation}
  a_0= -{9.30\times 10^{-5}
    {(1-0.35\chi_{\rm He})}^2{(1-0.74\chi_{\rm He})}^{1/3}
    \chi_{\rm He}^{8/3}
    ({\rm Mpc} \, k)^2 H^3
    \over
    (1-0.60\chi_{\rm He}+0.23{\chi_{\rm He}}^2)
    (1-0.24\chi_{\rm He})^{1/6}
    (1-0.29\chi_{\rm He})
    (1-0.33\chi_{\rm He})^{1/6}
    (1-0.77\chi_{\rm He})
    \tau_{24}^{1/6}}
\end{equation}
This is largest (in magnitude) at $\chi_{\rm He}=1$, where it is
\begin{equation}
  a_0= -2.79\times 10^{-4}{
    ({\rm Mpc} \, k)^2 H^3
    \over
    \tau_{24}^{1/6}}
\end{equation}

For a sufficiently small time, $a_0$ will be effectively constant and
the growth rate will be
\begin{equation}
  \label{growth1}
  \omega_g = \sqrt[3]{-a_0}
\end{equation}
We will show that there is no growth for any such small time interval,
so there will be no growth for any larger time interval.

\subsection{Range of the Instability}

\label{range_sec}

The minimum scale for this instability is the point where the
assumption that the medium is optically thick breaks down;
it is the optical path length (which should be the minimum of the lengths for
a helium or a hydrogen ionization, but for convenience we take it to be the
length for a helium ionization):
\begin{equation}
l_{\rm min}={1\over\sigma_{\rm He}(1-\chi_{\rm He})n_{\rm He}}
\end{equation}
or (still with $E_0=E_1$)
\begin{equation}
l_{\rm min}z'=5.46{\rm Mpc}{1\over(1-\chi_{\rm He}){z'}^2}
\end{equation}
or in terms of wave number,
\begin{equation}
  \label{MFP-lim}
  k_{\rm max}={2 \pi \over l_{\rm min}z'} =
    1.15{\rm Mpc}^{-1}(1-\chi_{\rm He}){z'}^2.
\end{equation}

The maximum scale comes from requiring that the
growth rate be larger than the expansion rate: $\omega_g > H$.
Taking $\omega_g$ from eq.~\ref{growth1} gives
\begin{eqnarray}
  & &k_{\rm min} = 104.\,{\rm Mpc}^{-1} \tau_{24}^{1/12}\times \nonumber\\
  & &\sqrt {
    (1-0.60\chi_{\rm He}+0.27{\chi_{\rm He}}^2)
    (1-0.24\chi_{\rm He})^{1/6}
    (1-0.30\chi_{\rm He})
    (1-0.33\chi_{\rm He})^{1/6}
    (1-0.77\chi_{\rm He})
    \over
    {(1-0.35\chi_{\rm He})}^2{(1-0.74\chi_{\rm He})}^{1/3}
    \chi_{\rm He}^{8/3}}
\end{eqnarray}
where the ionization is related to the redshift through
eq.~\ref{tau-z}.

\begin{figure}
\ifnum\outtype=1
 \epsffile{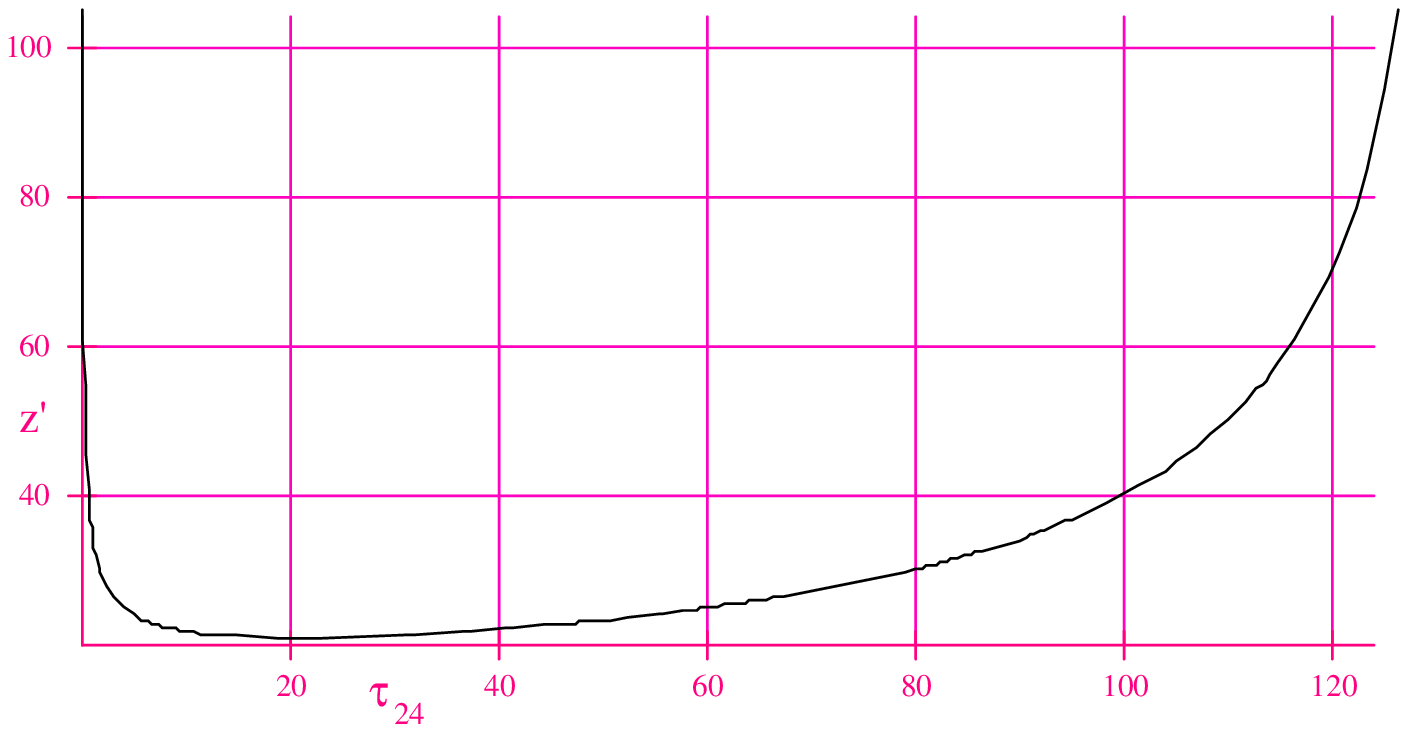}
\fi
\caption{Minimum $z'=z+1$ vs $\tau_{24}=\tau_\nu/10^{24}$.
For small $\tau_{24}$ (large photon production rate), the minimum
redshift is large because the large equilibrium ionization leads to a large
mean free path. 
For large $\tau_{24}$ the minimum redshift is large because the low
ionization leads to low instability growth.}
  \label{minz-tau}
\end{figure}

Requiring that the instability have a non-zero range of scales, or
\begin{equation}
k_{\rm min} <  k_{\rm max}
\end{equation}
gives a limit on $\chi_{\rm He}$ or equivalently $z'$ for a particular
$\tau_{24}$.
Figure~\ref{minz-tau}
shows the lower bound on $z'$ as a function of $\tau_{24}$.
The lowest $z'$ at which
there is a finite range for the instability is
$z'=21.0$ (for $\tau_{24} = 20.7$ and $\chi_{\rm He}=0.665$).

An additional constraint is that the assumption of ionization
equilibrium be valid.  Define the ionization rate per helium ion as
$\omega_{\rm Rec} = \alpha_{\rm He} n_e = \alpha_{\rm He} \chi \rho$.
For these parameters $\omega_{\rm Rec} = 1.01 H$ so equilibrium still
(marginally) holds.

\subsection{Compton Cooling}

In the post-recombination, pre-galactic universe of our model,
the dominant cooling mechanism is Compton scattering
off the background radiation (since in the DDM model
there will always be a significant ionized fraction).

The power lost to Compton cooling is
\begin{equation}
{\cal L}
   = 4 a T_\gamma^4{\sigma_{\rm Thompson} \over c m_{\rm electron}}
    n_e(T-T_\gamma)
   = 4 a T_\gamma^4{\sigma_{\rm Thompson} \over c m_{\rm electron}}
     ({\chi u \over 1+\chi} - n \chi T_\gamma)
\end{equation}
which consists of a cooling term proportional to
 $u {\chi \over 1 + \chi}$
and a heating term proportional to 
 $n \chi$.
The cooling term increases with ionization and thus opposes any
instability.  The heating term is approximately constant with density
or ionization
(since $\chi$ varies approximately as $1/n$)
but is also (weakly) stabilizing.

\begin{figure}
\ifnum\outtype=1
 \epsffile{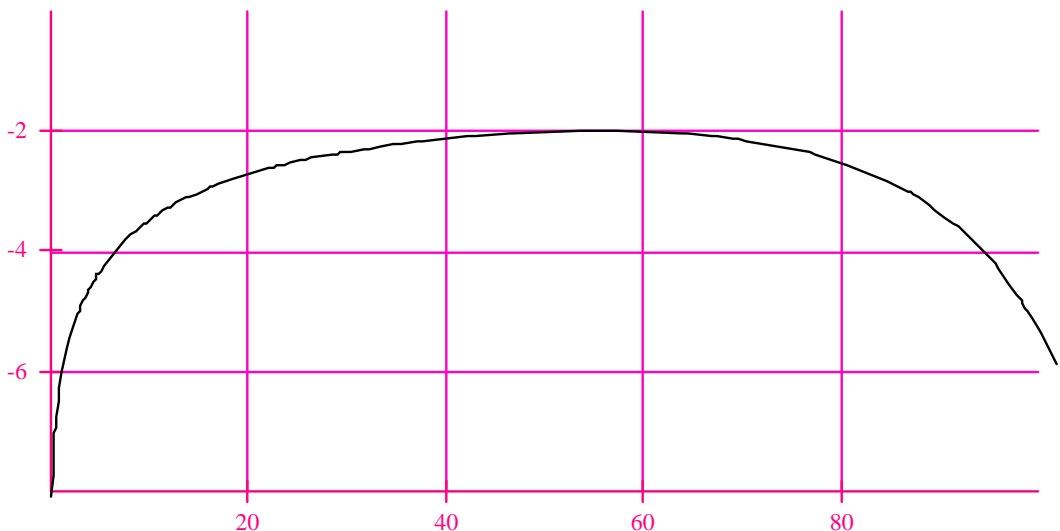}
\fi
\caption{${{\cal L}_n \over {\cal P}_n}$ vs $\tau_{24}$.
The curve follows the inverse of the minimum $z'$ curve 
(since Compton cooling is much more effective at large $z$),
with a shift towards
larger $\tau_{24}$ (where the ionization and thus the cooling is less).}
  \label{w-tau}
\end{figure}

Since Compton cooling depends more directly on the ionization
(${\cal L}_\chi \approx {\cal L}$),
its stabilizing effect dominates if the cooling is comparable in
magnitude to the photo-ionization heating.
With a simple model for the temperature\footnote{
To guarantee that the model temperature was a lower bound, we took the
ionization in ${\cal L}$ to be constant at 1 and the ionization in
${\cal P}$ to be constant at 0 and we approximated the resulting
integral for $u$ in a way that gives about half the actual value for the
difference in the matter temperature and the photon temperature.
}, we calculate that at the lowest redshift
possible, $z'=21.0$,
the ratio of the cooling density dependence to the heating density
dependence is
\begin{equation}
{{\cal L}_n \over {\cal P}_n}<-2.68.
\end{equation}
Figure~\ref{w-tau} 
shows ${{\cal L}_n \over {\cal P}_n}$ as a function of
$\tau_{24}$ at the minimum $z'$ using the same thermal model.
It's smallest magnitude is
\begin{equation}
{{\cal L}_n \over {\cal P}_n}<-2.00
\end{equation}
at $z'=24.3$, $\tau_{24}=55.7$, and $\chi_{\rm He}=0.409$.\footnote{
For these parameters, $\omega_{\rm Rec} = 0.612 H$, so the assumption of
ionization equilibrium is invalid.
Enforcing that assumption would require an even less favorable ratio.}

Thus, the stabilization of Compton cooling will always dominate
the destabilizing effects of the heating.

\section{Scenario in which Recombination is Ignored}

At redshifts or ionization rates  which are low enough
that Compton cooling is not
significant, recombination is also less important.
We model this regime as having no recombination, which strengthens the
instability, but reduces the time over which it can operate.



\subsection{Ionization Equations}

We solve for the ionization by still assuming
photon production to be in equilibrium with ionization:
\begin{equation}
  \Re=(\sigma_{\rm He 0}n_{\rm He\,I}+\sigma_{\rm H 0}n_{\rm H\,I})n_{\gamma 0}
  \label{photon2}
\end{equation}
but ionization is not in equilibrium with recombination:
\begin{equation}
  {d \chi_{\rm He} \over d t} =
    n_{\gamma 0}\sigma_{\rm He 0}(1-\chi_{\rm He})
\label{chi_He_dot}
\end{equation}
\begin{equation}
  {d \chi_{\rm H} \over d t} =
    n_{\gamma 0}\sigma_{\rm H 0}(1-\chi_{\rm H}).
\label{chi_H_dot}
\end{equation}

Taking the ratio of eqs.~(\ref{chi_He_dot}) and (\ref{chi_H_dot}) we get
\begin{equation}
  {d \chi_{\rm H} \over d \chi_{\rm He}} =
    {\sigma_{\rm H 0} \over \sigma_{\rm He 0}}\,
    {1-\chi_{\rm H} \over 1-\chi_{\rm He}}
\end{equation}
or
\begin{equation}
  1-\chi_{\rm H} =
    c_1 (1-\chi_{\rm He})^{\sigma_{\rm H 0} \over \sigma_{\rm He 0}}
\label{chi_H_chi_He}
\end{equation}
where $c_1$ is determined by the initial conditions.  This is an exact
equation which holds for the background ionization and which can be
varied to yield a relationship between the hydrogen and helium
perturbations:
\begin{equation}
  \label{dHe-dH}
  \delta \chi_{\rm He} =
  c_1 {\sigma_{\rm H 0} \over \sigma_{\rm He 0}}
  (1-\chi_{\rm He})^{{\sigma_{\rm H 0} \over \sigma_{\rm He 0}}-1}
  \delta \chi_{\rm H}
\end{equation}

From eqs.~(\ref{chi_He_dot}), (\ref{chi_H_chi_He}), and (\ref{photon2}) we get
\begin{equation}
  (1-\mu) (1-\chi_{\rm H}) + \mu (1-\chi_{\rm He}) =
  -\int {\Re \over n} d t
\label{chi_He}
\end{equation}

Note that ${\Re \over n}$ is a constant in the unperturbed
background, so for background quantities:
\begin{equation}
  (1-\mu) (1-\chi_{\rm H}) + \mu (1-\chi_{\rm He}) =
  -{\Re \over n} t
\label{chi_He_0}
\end{equation}
which gives the ionization as an implicit function of time or (as we
shall use it) gives the time as a function of the background ionization.

Using the following formula (valid for a flat, matter-dominated
universe) we can express the redshift in terms of time and thus
ionization.
\begin{equation}
t = {2 \over 3} {1 \over H_0} (z')^{-{3 \over 2}}
\end{equation}

Varying eq.~\ref{chi_He} gives
\begin{equation}
  \label{d-chi_He}
  (1-\mu) \delta \chi_{\rm H} + \mu \delta \chi_{\rm He} =
  -{\Re \over n} \int \delta \ln n d t
\end{equation}
which relates the variation of the density to the variation of the
ionization.

\subsection{The Variation of Heating with Density}

The heating function, ${\cal P}$, is
simpler than in the first scenario
because there are no photons at reprocessed energy levels:
\begin{equation}
  {\cal P} = n_{\gamma 0}[(E_0-{\rm Ry})\sigma_{\rm H 0}n_{\rm H\,I}+
   (E_0-E_1)\sigma_{\rm He 0}n_{\rm He\,I}].
  \label{P2}
\end{equation}

Using eq.~\ref{photon2} to eliminate $n_{\gamma 0}$ gives
\begin{equation}
  {\cal P} = {\cal R} \left[(E_0-{\rm Ry}) -
    (E_1-{\rm Ry}) {1 \over 1 +
      {1-\mu \over \mu}
      {\sigma_{\rm H 0} \over \sigma_{\rm He 0}}
      {1-\chi_{\rm H} \over 1-\chi_{\rm He}}}
    \right]
  \label{P3}
\end{equation}

As in the equilibrium case, the heating function depends on the
density only through the ionization.  Varying ${\cal P}$ with respect
to the ionization and using the relationship between ionization and
density variations (eq.~\ref{d-chi_He}) gives $\delta {\cal P}_n$ which
substituted into eq.~\ref{norec-1} gives
\begin{equation}
  \label{norec-2}
  {\partial \over \partial t} {\partial \over \partial t}
  {\partial \over \partial t} \delta \ln n +
  a_2
  {\partial \over \partial t} {\partial \over \partial t} \delta \ln n +
  a_1
  {\partial \over \partial t} \delta \ln n -
  a_0 \int \delta \ln n dt
\end{equation}
where
\begin{equation}
  a_0={(\gamma-1)(1-\mu)\over(3\mu+1)\mu^2}
  {\sigma_{\rm H 0} \over \sigma_{\rm He 0}}
  \left(1- {\sigma_{\rm H 0} \over \sigma_{\rm He 0}}\right)
  {E_1-{\rm Ry} \over m_{\rm proton}}
  {1-\chi_{\rm H} \over (1-\chi_{\rm He})^2}
  {1 \over \left(1 +
      {1-\mu \over \mu}
      {\sigma_{\rm H 0} \over \sigma_{\rm He 0}}
      {1-\chi_{\rm H} \over 1-\chi_{\rm He}}\right)^3}
  {{\cal R} \over n}^2 k^2 z'^2
\end{equation}
and $a_1$ and $a_2$ are given by eqs.~\ref{a_1} and \ref{a_2}.

\subsection{Bounds to the Growth}

In this scenario an instability can grow, though the total growth is
less than one e-folding so it is insignificant.

\begin{figure}
\ifnum\outtype=1
 \epsffile{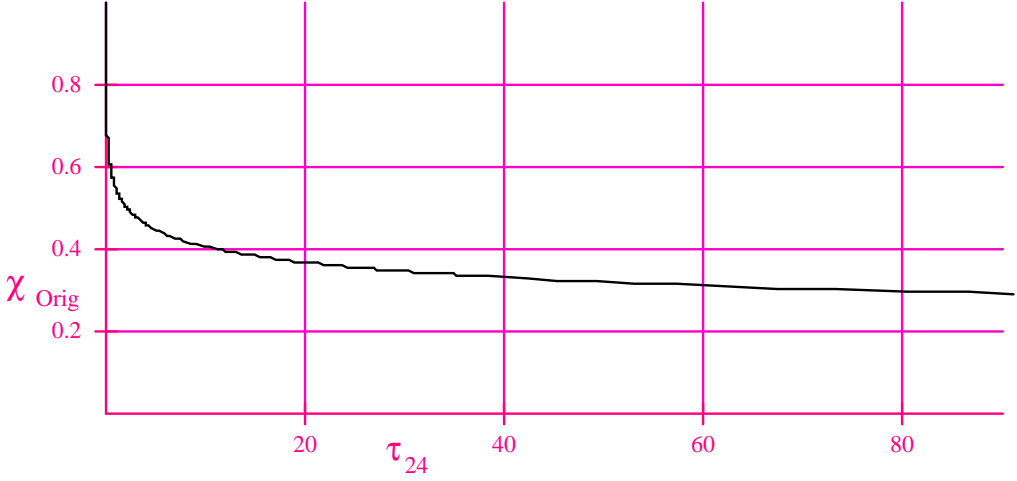}
\fi
\caption{Starting $\chi_{\rm He}$ for no-recombination
 conditions vs $\tau_{24}$.  This is the recombination-equilibrium
 ionization where the Compton cooling ceases to dominate the heating.
 The steep slope near $\tau_{24}=0$ reflects the fact that the heating
 approaches zero as $\chi_{\rm He}$ goes to one, so there needs to be
 a high photon production rate to compensate.}
 \label{norec-chi-tau}
\end{figure}

The time available for growth is bounded by the neutral atoms
being used up, since there is no recombination.
The starting ionization will be greater than zero since some
ionization will occur while Compton cooling damps any growth.
We take $\chi_{\rm Orig}$ to be the recombination-equilibrium
helium ionization (eqs.~\ref{n_vs_chi} and~\ref{chi_h1})
at the red-shift where the damping from the Compton
cooling just becomes weaker than the instability growth.  This is
generous to the model in three ways: 1) assuming recombination
equilibrium always over estimates the remaining neutral fraction; 2)
the Compton cooling is calculated for matter at the photon background
temperature, where it is lowest (the net cooling will be zero at this
temperature, but there is a non-zero derivative of the
cooling with respect to helium ionization); and 3) the growth will actually
be zero at that point, not its full value without Compton cooling.
Figure~\ref{norec-chi-tau} shows $\chi_{\rm Orig}$
as a function of $\tau_{24}$.
$z'_{\rm Orig}$ and the original $\chi_{\rm H}$ can be obtained from
$\chi_{\rm Orig}$ and $\tau_{24}$ using eqs.~\ref{tau-z} and
\ref{chi_h1}.
Those parameters then determine $c_1$ from eq.~\ref{chi_H_chi_He}
and the constant of integration in
eq.~\ref{d-chi_He}.

The largest ionization,
$\chi_{\rm Final}$ is the value of the helium ionization at which the
mean-free-path exceeds the scale under consideration (i.e. it depends
on $k$).  It is determined for a given $k$ from eq.~\ref{MFP-lim}
(given the redshift/ionization relationship of eq.~\ref{d-chi_He}).

To determine the total growth,
we broke the evolution down in
steps with $a_0$ taken to be constant at its maximum value in each
step (the maximum is achieved for the smaller ionization at the
start of the step).
For constant $a_0$ (and ignoring $a_2$ and $a_3$, which only damp any
growth) the
growing solution is $\delta \ln n \propto \exp(\sqrt[4]{a_0} t)$.
We took steps in $\chi_{\rm He}$ instead of time since we have time
and redshift as a function of $\chi_{\rm He}$ and not vice-versa.
Thus the growth in $\delta \ln n$ is bounded by
$\exp(\sum_{i=0}^{N-1}(t(\chi_{i+1}) -
t(\chi_{i}))\sqrt[4]{a_0(\chi_{i}})$ for any $N$ (with a tighter bound
for a larger $N$) with
$\chi_{i}= \chi_{\rm Orig} + {i \over N} (\chi_{\rm Final} -
\chi_{\rm Orig})$.

\begin{figure}
\ifnum\outtype=1
 \epsffile{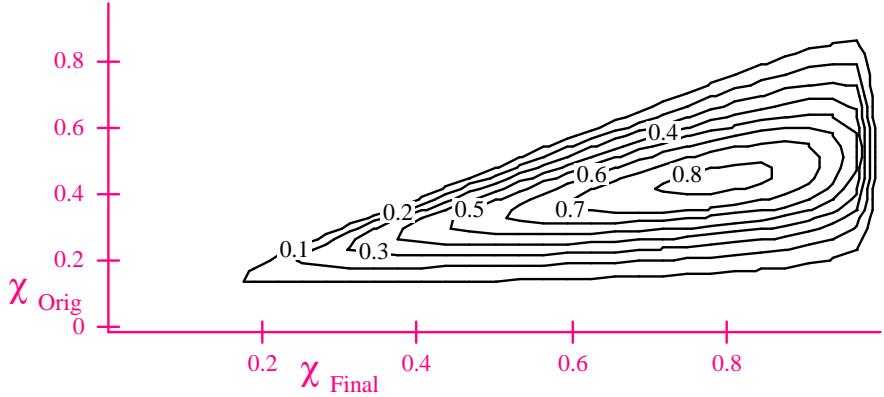}
\fi
\caption{Logarithm of the total growth factor vs the initial and final
  ionization for no-recombination growth.
  Since ionization increases with time, we clearly need the final
  ionization to be larger than the initial ionization, with larger
  values of their difference giving more time for growth.
  However, very small values for the original ionization correspond to
  low photon production rates, which give small overall growth.
  Large values for the final ionization correspond to large spatial
  scales, where the growth is also reduced.}
  \label{norec-growth}
\end{figure}

Figure~\ref{norec-growth} shows the log of the total growth (for $N=40$) as a function of
$\chi_{\rm Orig}$ and $\chi_{\rm Final}$.
It's largest value is $\exp(.83)$ at $\chi_{\rm Orig}=.44$ and 
$\chi_{\rm Final}=.79$ (for these parameters the model is actually
invalid because recombination is large, but it shows that within the
space of validity the growth will be even less).

\section{Conclusions}

Though there is an interesting instability involving helium ionization
by line radiation, it doesn't seem to have a cosmological application
in the DDM scenario.
As other candidate radiation sources presume the existence of
small-scale structure, we conclude that this class of instability is
unlikely to play a role in the initial formation of structure from
smooth cosmic gas.

\section{Acknowledgments}
\acknowledgements
This work was supported by
NASA grants NAGW 2569, NAGW 2523, and NAG 5 2793
at the University of Washington.
We are grateful to the referee, Dr. A. Meiksin, for
for his careful review and for several helpful
suggestions.

\appendix

\section{Calculation of Two-Photon Decay Parameters}

\begin{figure}
\ifnum\outtype=1
  \epsfysize=3in \epsffile[-20 0 210 300]{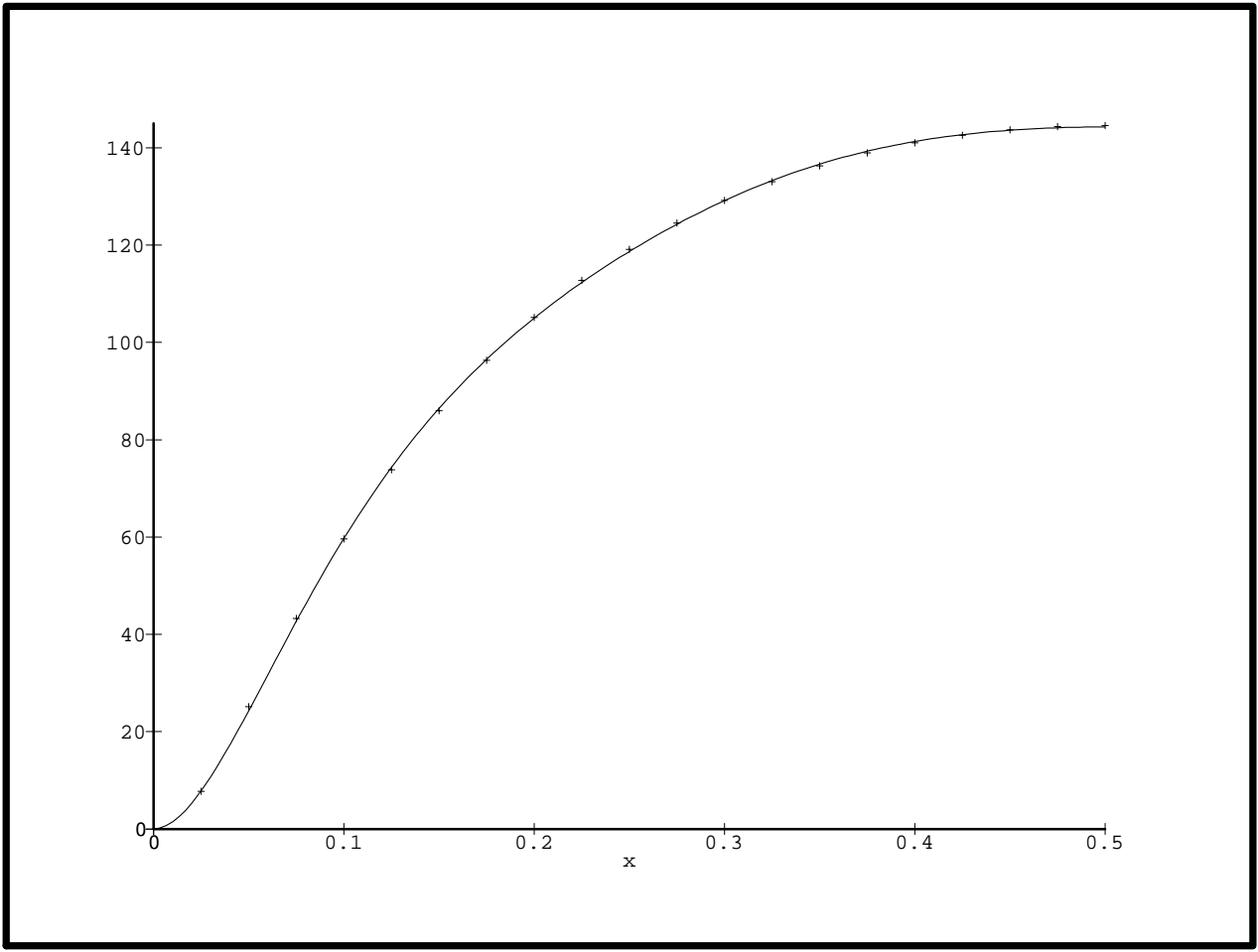}
\fi
\caption{He $2 {}\,^1\!S$ two-photon decay rate versus energy
  ($x=\nu/E_4$).
  The fitting formula is
  $A(x)= -0.98 p(2,x) +2.83 p(3,x) -4.26 p(4,x) +2.15 p(5,x)$,
  where $p(j,x)=300((2x-1)^{2j}-1-j ((2x-1)^2-1))$
  (this form is chosen so that each
  term is symmetric around x=0.5 and is zero and has zero
  slope at $x=0$ and $x=1$).
  The RMS error of the fit is 0.34.
}
  \label{2-photon-fit}
\end{figure}

The He $2 {}\,^1\!S$ state
decays by the emission of two photons whose energy sums to $E_4$.
The strength of the transition which emits one of the photons in the
range $\nu$ to $\nu+d\nu$ is $A(\nu)d\nu$.
We did an empirical fit to the average of the two tables of values for
$A$ given in \citeN{two-photon}.
Figure~\ref{2-photon-fit} shows the fit and compares it to
Jacobs's values.

The total strength for emission of two photons is 
$A_{\rm Tot}\equiv\int_0^{E_4}A(\nu)d\nu$.
The total strength for the events in
which a particular one of the photons is energetic enough to ionize hydrogen is
$A_{\rm Ry}\equiv\int_{\rm Ry}^{E_4}A(\nu)d\nu$.
Since we have a hydrogen
ionization if either photon is energetic enough, the proportion of
decays which ionize hydrogen is\footnote{This differs slightly from .56 given
  in \citeN{Osterbrock:1}, perhaps because his value was
  not updated when he updated $E_4$ from 20.7 to 20.6.}

\begin{equation}
  p_1=2{A_{\rm Ry} \over A_{\rm Tot}}=0.554
  \label{p1}
\end{equation}

Weighting the strength of decay by the energy of the decay gives the
average energy of emission, here taken over only those photons which
can ionize hydrogen:
\begin{equation}
  E_5={\int_{\rm Ry}^{E_4}\nu A(\nu) d\nu  \over A_{\rm Ry}}= 16.1 {\rm eV}
\label{E5}
\end{equation}

Since the absorption cross section depends on frequency, it should also be
calculated as a weighted average:
\begin{eqnarray}
\sigma_{\rm H 5}&=&{\int_{\rm Ry}^{E_4}  \sigma_{\rm H}(\nu) A(\nu) d\nu  \over A_{\rm Ry}}\\
&=&1.050 \sigma_{\rm H}(E_5)
\end{eqnarray}
so the weighting gives a 5\% correction compared to using the
absorption cross section for line radiation at energy $E_5$.

\section{Ignored Effects}

{\bf Recombinational cooling}, or the thermal energy lost when two
particles which both have average kinetic energy $T$ combine to one
particle with average energy $T$, is insignificant.  Since each
ionization results in a recombination in the first scenario, the ratio
of recombination cooling to ionization heating is roughly the ratio of
the average energy deposited by an ionization to the temperature, or
about $T/6.5$eV.  Compton cooling keeps the matter temperature
comparable to the radiation temperature, which is much less than 1 eV.
For the second scenario there is no recombination, and thus no
recombinational cooling.  Note that no collisional ionizations happen
at the low temperatures considered here.

{\bf Collisional Transitions} between helium levels are
insignificant at the cosmological densities which are of primary
concern in this analysis.  The critical density given in
\citeN{Osterbrock:1} above which one must consider
collisional transitions is $\sim 5000 cm^{-3}$, much larger than
cosmological densities of $\sim 10^{-7} z'^3 cm^{-3} < 100 cm^{-3}$
for $z' < 1000$.

The
magnitude of the {\bf photon redshift} over one mean-free-path (for,
e.g., an $E_1$ photon to ionize helium)
is
\begin{equation}
\Delta z = 1.8\times{10}^{-3}h{1 \over
  z'\,^{3 \over 2} (1-\chi_{\rm He})}
\end{equation}
An $E_1$ photon will be redshifted by more than one helium thermal
Doppler over one mean-free-path for $z \lesssim 100$; this
effectively reduces the absorption cross section of helium, which
weakens the instability.
The redshift over the mean-free-path of a one Rydberg photon is
greater than a 
hydrogen thermal Doppler width for all $z \lesssim 50$; for
smaller redshifts than that, some hydrogen recombinations direct to
ground will be allowed, which slightly strengthens the instability, as
can be seen by considering eq.~(\ref{growth-condition}) with
increased $\alpha_{{\rm H} 2}$.

A {\bf distribution of initial photon energies}
will result in reduced growth over a line spectrum with an energy just
above the Helium ionization threshold simply because any photons much above
the threshold are less effective
and any photons below the threshold actually oppose the instability.

{\bf Lyman-alpha trapping} will occur, since for $z \gtrsim 40$, the
redshifting in one mean-free-path is insufficient to bring a
Lyman-$\alpha$ photon more than one thermal Doppler width away from
the line center, but because the time photons spend traveling between
encounters is much larger than the time a hydrogen atom spends in the
${}\,^2\!P$ state, only approximately $10^{-7}$ of the neutral hydrogen
atoms are excited, and this
does not affect the instability.

{\bf Photon diffusion} (relaxing the assumption that the medium is
optically thick) opposes an instability, since the 
more-ionized/less-dense regions are more optically thin and would thus lose
photons (and therefor energy) to the less-ionized/more-dense regions.

{\bf Thermal conduction} opposes any instability and its
effects are insignificant on the large scales considered here.

{\bf Collisional ionization} goes up with temperature, but it goes up with
density squared so it would also act against an instability.
Its rate is insignificant at the temperatures and densities
of the post-recombination universe.

The {\bf temperature dependence of the recombination coefficients} does
not have a large effect because only the ratios of recombination
coefficients enter the formulas, and the recombination coefficients for
helium and hydrogen both depend on temperature to roughly the same
power ($3/4$ for $T \approx 10000 K$).

Any {\bf delay in thermalization} has no effect, since Compton
cooling and pressure are both linear in the sum of the energy of the
particles, so the distribution of the energy does not matter.

\bibliographystyle{apj}
\bibliography{apjmnemonic,../cosmo}

\newpage

\ifnum\outtype=1
  \newpage
\begin{deluxetable}{|ll|}
  \tablewidth{0pt}
  \tablecaption{Symbols}
  \tablehead{\colhead{Symbol/Value}&\colhead{Explanation}}
  \startdata

  \parbox[t]{1.6 in}{
    $h$
    }&
  \parbox[t]{3.6 in}{
    The Hubble Parameter in units of 100 km/s/Mpc.
    }\nl

  \parbox[t]{1.6 in}{
    $\omega_g =-i \omega$
    }&
  \parbox[t]{3.6 in}{
    Instability growth rate.
    }\nl

  \parbox[t]{1.6 in}{
    $\vec{k}$
    }&
  \parbox[t]{3.6 in}{
    Wave vector in Fourier decomposition.
    }\nl

  \parbox[t]{1.6 in}{
    $k\equiv|k|$
    }&
  \parbox[t]{3.6 in}{
    Wave number (inverse of size scale).
    }\nl

  \parbox[t]{1.6 in}{
    $n$
    }&
  \parbox[t]{3.6 in}{
    Total spatial density of hydrogen plus helium nuclei
    (neutral and ionized).
    }\nl

  \parbox[t]{1.6 in}{
    $u={1 \over \gamma -1}(1+\chi)n T_m$
    }&
  \parbox[t]{3.6 in}{
    Internal thermal energy of the gas.
    }\nl

  \parbox[t]{1.6 in}{
    $\gamma = 5/3$
    }&
  \parbox[t]{3.6 in}{
    Adiabatic expansion coefficient.
    }\nl

  \parbox[t]{1.6 in}{
    ${\cal P}$
    }&
  \parbox[t]{3.6 in}{
    Heat input per unit volume.
    }\nl

  \parbox[t]{1.6 in}{
    $v_s$
    }&
  \parbox[t]{3.6 in}{
    The unperturbed speed of sound.
    }\nl

  \parbox[t]{1.6 in}{
    $\mu={Y_p \over 4 - 3 Y_p} \approx 0.073$
    }&
  \parbox[t]{3.6 in}{
    Fraction by number of helium atoms ($Y_p \approx .24$ is used here).
    }\nl

  \parbox[t]{1.6 in}{
    $\omega_u = {\cal P}_u\\
     \omega_k = v_s k\\
     \omega_n = -{{\cal P}_n n \over\gamma u}$
    }&
  \parbox[t]{3.6 in}{
    Miscelaeous rates.
    }\nl

  \parbox[t]{1.6 in}{
    $p_1=0.554$
    }&
  \parbox[t]{3.6 in}{
    Number of photons produced in a two-photon decay of the He $2 {}\,^1\!S$ state
    which can ionize H.
    }\nl

  \parbox[t]{1.6 in}{
    $n_{\gamma 0}, n_{\gamma 1}$, etc.}&
  \parbox[t]{3.6 in}{
    Spatial density 
    of photons with energy $E_0, E_1, $ etc.
    }\nl

  \parbox[t]{1.6 in}{
    $n_{\rm He\,II}=n\mu \chi_{\rm He}$
    }&
  \parbox[t]{3.6 in}{
    Number density of ionized helium.
    }\nl

  \parbox[t]{1.6 in}{
    $n_{\rm He\,I}=n\mu(1- \chi_{\rm He})$
    }&
  \parbox[t]{3.6 in}{
    Number density of neutral helium.
    }\nl

  \parbox[t]{1.6 in}{
    $n_{\rm H\,II}=n(1-\mu) \chi_{\rm H}$
    }&
  \parbox[t]{3.6 in}{
    Number density of ionized hydrogen.
    }\nl

  \parbox[t]{1.6 in}{
    $n_{\rm H\,I}=n(1-\mu)(1- \chi_{\rm H})$
    }&
  \parbox[t]{3.6 in}{
    Number density of neutral hydrogen.
    }\nl

  \parbox[t]{1.6 in}{
    $n_e=n_{\rm He\,II}+n_{\rm H\,II}=n \chi$
    }&
  \parbox[t]{3.6 in}{
    Number density of electrons.
    }\nl

  \parbox[t]{1.6 in}{
    $\chi_{\rm He}$
    }&
  \parbox[t]{3.6 in}{
    Fractional ionization of helium.
    }\nl

  \parbox[t]{1.6 in}{
    $\chi_{\rm H}$
    }&
  \parbox[t]{3.6 in}{
    Fractional ionization of hydrogen.
    }\nl

  \parbox[t]{1.6 in}{
    $\chi = \mu  \chi_{\rm He}+(1-\mu) \chi_{\rm H}$
    }&
  \parbox[t]{3.6 in}{
    Fractional ionization of all species combined.\\
    }\nl

  \parbox[t]{1.6 in}{
  $\sigma_{\rm H}(E)=6.30\times{10}^{-18}\times\\
  \;\;x^{-4} {e^{4(1-y \, {\rm arccot}(y))} \over 1-e^{-2\pi y}}{\rm cm}^2\\
  \mbox{\hspace{10pt}}x={E \over {\rm Ry}}\\
  \mbox{\hspace{10pt}}y={1 \over \sqrt{x-1}}$
  }&
  \parbox[t]{3.6 in}{
    Absorption cross section of hydrogen for
    photons with energy $E$.\tablenotemark{b}
    }\nl

\ifnum\outtype=1
\tablebreak
\fi

  \parbox[t]{1.6 in}{
  $\sigma_{\rm H \, \mit m}=\sigma_{\rm H}(E_m)$
  }&
  \parbox[t]{3.6 in}{
    Absorption cross section of hydrogen for
    photons with energy $E_m$, for some $m$.
    }\nl

  \parbox[t]{1.6 in}{
    $\sigma_{\rm He \, \mit m}=\sigma_{\rm H \, \mit m} \times \\
    \;\;(6.53{E_m \over E_1}-0.22)$
  }&
  \parbox[t]{3.6 in}{
   Absorption cross section of helium for
   photons with energy $E_m$, for some $m$.\tablenotemark{c}
   }\nl

  \parbox[t]{1.6 in}{
    $\alpha_{{\rm H} 2}=2.59\times{10}^{-13}{{cm}^3 \over s}$
    }&
  \parbox[t]{3.6 in}{
    Recombination coeff.\ for H to
    all excited levels.\tablenotemark{a}
    }\nl

  \parbox[t]{1.6 in}{
    ${\alpha}_{{\rm He} 2}=2.73\times{10}^{-13}{{cm}^3 \over s}$
    }&
  \parbox[t]{3.6 in}{
    Recombination coeff.\ for He to
    all excited levels.\tablenotemark{a}
    }\nl

  \parbox[t]{1.6 in}{
    ${\alpha}_{{\rm He} 2 {}\,^1\!P} = {2 \over 3}
    \left({1 \over 4}\alpha_{{\rm He} 2}\right)$
    }&
  \parbox[t]{3.6 in}{
    Recombination coeff.\ for He to
    excited singlet levels which cascade to the $2{}^1\!P$ state.
    }\nl

  \parbox[t]{1.6 in}{
    ${\alpha}_{{\rm He} 2 {}\,^1\!S} = {1 \over 3}
    \left({1 \over 4}\alpha_{{\rm He} 2}\right)$
    }&
  \parbox[t]{3.6 in}{
    Recombination coeff.\ for He to
    excited singlet levels which cascade to the $2 {}\,^1\!S$ state.
    }\nl

  \parbox[t]{1.6 in}{
    ${\alpha}_{{\rm He} 2 {}\,^3\!P} = {3 \over 4}\alpha_{{\rm He} 2}$
    }&
  \parbox[t]{3.6 in}{
    Recombination coeff.\ for He to all
    excited triplet levels (all of which lead to the $2 {}\,^3\!P$ state).
    }\nl

  \parbox[t]{1.6 in}{
    ${\alpha}_{\rm He 1}=1.59\times{10}^{-13}{{cm}^3 \over s}$
    }&
  \parbox[t]{3.6 in}{
    Recombination coeff.\ for He to
    ground.\tablenotemark{a}
    }\nl

  \parbox[t]{1.6 in}{
    $\Re = n_\nu / \tau_\nu$
    }&
  \parbox[t]{3.6 in}{
    Rate of production of primary photons.
    }\nl

  \parbox[t]{1.6 in}{
    $\delta = {E_0 \over E_1} - 1$
    }&
  \parbox[t]{3.6 in}{
    Dimensionless excess energy of ionization photon.
    }\nl

  \parbox[t]{1.6 in}{
    $\phi={T_m \over T_\gamma}$
    }&
  \parbox[t]{3.6 in}{
    Ratio of the matter temperature to the radiation temperature.
    }\nl

  \parbox[t]{1.6 in}{
    $z'=1+z$}&
  \parbox[t]{3.6 in}{
    Inverse of the cosmic scale factor.
    }\\

  \parbox[t]{1.6 in}{
    $\tau_{24} = \tau_\nu / 10^{24} {\rm sec}$
    }&
  \parbox[t]{3.6 in}{
    Neutrino decay rate in units of $10^{24}$ seconds.
    }\nl
\enddata

\tablenotetext{a}{From Osterbrock 1989 
 \nocite{Osterbrock:1} for $T=10000K$.}
\tablenotetext{b}{From Spitzer 1978\nocite{Spitzer}.}
\tablenotetext{c}{From Brown 1971\nocite{Brown:photoionization}.}
\end{deluxetable}

\begin{deluxetable}{|lll|}
  \tablewidth{0pt}
  \tablecaption{Energy Values}
  \tablehead{\colhead{Sym.}&\colhead{Value}&\colhead{Explanation}}
  \startdata
  $E_0$&\nodata&Primary photon energy\\
  $E_1$&24.6 eV\tablenotemark{a}&Ionization energy of He\\
  $E_2$&19.8 eV\tablenotemark{a}&Energy of He $2 {}\,^3\!P$\\
  $E_3$&21.2 eV\tablenotemark{a}&Energy of He $2 {}\,^1\!P$\\
  $E_4$&20.6 eV\tablenotemark{a}&Energy of He $2 {}\,^1\!S$\\
  $E_5$&16.1 eV\tablenotemark{b}&Avg. energy of $\gamma$ absorbed by H in 2$\gamma$
  decay of He $2 {}\,^1\!S$\\
  ${\rm Ry}$&13.6 eV&H ionization energy\\
  $\Delta E$&$E_0-{\rm Ry}$&Energy deposited in a direct H ionization\\
  $\Delta E_{\rm He}$&$E_0-E_1$&Energy deposited in a He ionization\\
  \enddata
  \label{ETable}
  \tablenotetext{a}{From Osterbrock 1989. 
    \nocite{Osterbrock:1}}
  \tablenotetext{b}{See Appendix~C.}
\end{deluxetable}
\fi

\end{document}